\documentclass[10pt]{iopart}
\usepackage{svg}
\usepackage{bm}
\usepackage{citesort}
\usepackage{float}
\makeatletter
\@namedef{ver@amsmath.sty}{}
\makeatother
\usepackage{amstext}

\usepackage{graphicx}
\usepackage{dcolumn}
\begin{document}

\nocite{}\title{Machine learning and excited-state molecular dynamics}

\noindent\author{Julia Westermayr$^1$ and Philipp Marquetand$^{1,2,3}$}

\noindent\address{$^1$ University of Vienna, Faculty of Chemistry, Institute of Theoretical Chemistry, W\"ahringer Str. 17, 1090 Wien, Austria\\
$^2$ Vienna Research Platform on Accelerating Photoreaction Discovery, University of Vienna, W\"ahringer Str. 17, 1090 Wien, Austria\\
$^3$ University of Vienna, Faculty of Chemistry, Data Science @ Uni Vienna, W\"ahringer Str. 29, 1090 Wien, Austria}

\ead{philipp.marquetand@univie.ac.at}
\vspace{10pt}

\noindent{}\begin{abstract}
Machine learning is employed at an increasing rate in the research field of quantum chemistry. While the majority of approaches target the investigation of chemical systems in their electronic ground state, the inclusion of light into the processes leads to electronically excited states and gives rise to several new challenges. Here, we survey recent advances for excited-state dynamics based on machine learning. In doing so, we highlight successes, pitfalls, challenges and future avenues for machine learning approaches for light-induced molecular processes.
\end{abstract}
\noindent{\it Keywords:} machine learning, photodynamics, photochemistry, excited states, quantum chemistry, spin-orbit couplings, nonadiabatic couplings.
\maketitle
\ioptwocol
\section{\label{sec:level1}Introduction}

Photosynthesis, photovoltaics, the processes that enable our vision or photodamage of biologically relevant molecules, such as DNA or peptides -- they all have one thing in common: The underlying processes are governed by a manifold of excited states after the absorption of light~\cite{Cohen2004,Levine2007ARPC,Turro2009,Yarkony2012CR,Barbatti2014,Ibele2019MP,Nelson2020CR,Mai2020ACIE,Matsika2018CR,Lischka2018CR,Ghosh2018CR,Norman2018CR,Casanova2018CR,Hestand2018CR,Penfold2018CR,Vacher2018CR,Crespo-Otero2018CR,Gonzalez2020}. They can be studied experimentally via several techniques, such as UV/visible spectroscopy, transient absorption spectroscopy, photoionization spectroscopy or ultrafast electron diffraction~\cite{Harris1989,Cheuk-Yiu1991,Zewail1994,Brixner2005,Neves-Petersen2009BJ,Iqbal2010JPCL,Malis2012JACS,Kowalewski2017CR,Soorkia2019CR}.
However, experimental techniques are to some extent blind to the exact electronic mechanism of photo-induced reactions. In order to get a more comprehensive understanding, theoretical simulations can complement experimental findings and can provide explanations for observed reactions~\cite{Matsika2018CR}. For instance, simulated UV spectra can be used to unveil the states relevant for photodamage and -stability of molecules~\cite{Tajti2009C,Barbatti2011JCP,Lu2014,Ruckenbauer2016JCP,Manathunga2016JCTC,Nogueira2017CS,Mai2018M,Rauer2018MC,Zobel2018JCTC} and the temporal evolution of molecules can be studied via nonadiabatic molecular dynamics (NAMD) simulations~\cite{Barbatti2006MP,Curchod2013JCP,Akimov2015JACS,Schapiro2015JCC,Rauer2016JACS,Ruckenbauer2016SR,Mai2017CP,Mai2019CS,Horton2019JCP,Heim2020JPCL}. The latter gives access to different reaction channels, branching ratios, and excited-state lifetimes and will be the main topic of discussion here.

While experimental techniques require large and costly setups, theoretical simulations require high-performance computing facilities due to expensive electronic structure computations. Especially NAMD simulations are seriously limited by the underlying quantum chemical calculations, making long and experimentally relevant simulation times inaccessible with conventional \emph{ab initio} methods. 
The larger the molecule becomes, the more electronically excited states are involved in reactions and the more complex their interactions become. This leads to non-linearly increasing costs of quantum chemical calculations and a compromise between accuracy and computational efficiency is indispensable. Relying on such expensive \emph{ab initio} potentials, only a couple of picoseconds can be simulated and the exploration of rare reaction channels is restricted due to bad statistics~\cite{Mai2016JPCL,Mai2017CP,Crespo-Otero2018CR}.

\begin{figure}[ht]
    \centering
    \includegraphics[scale=0.44]{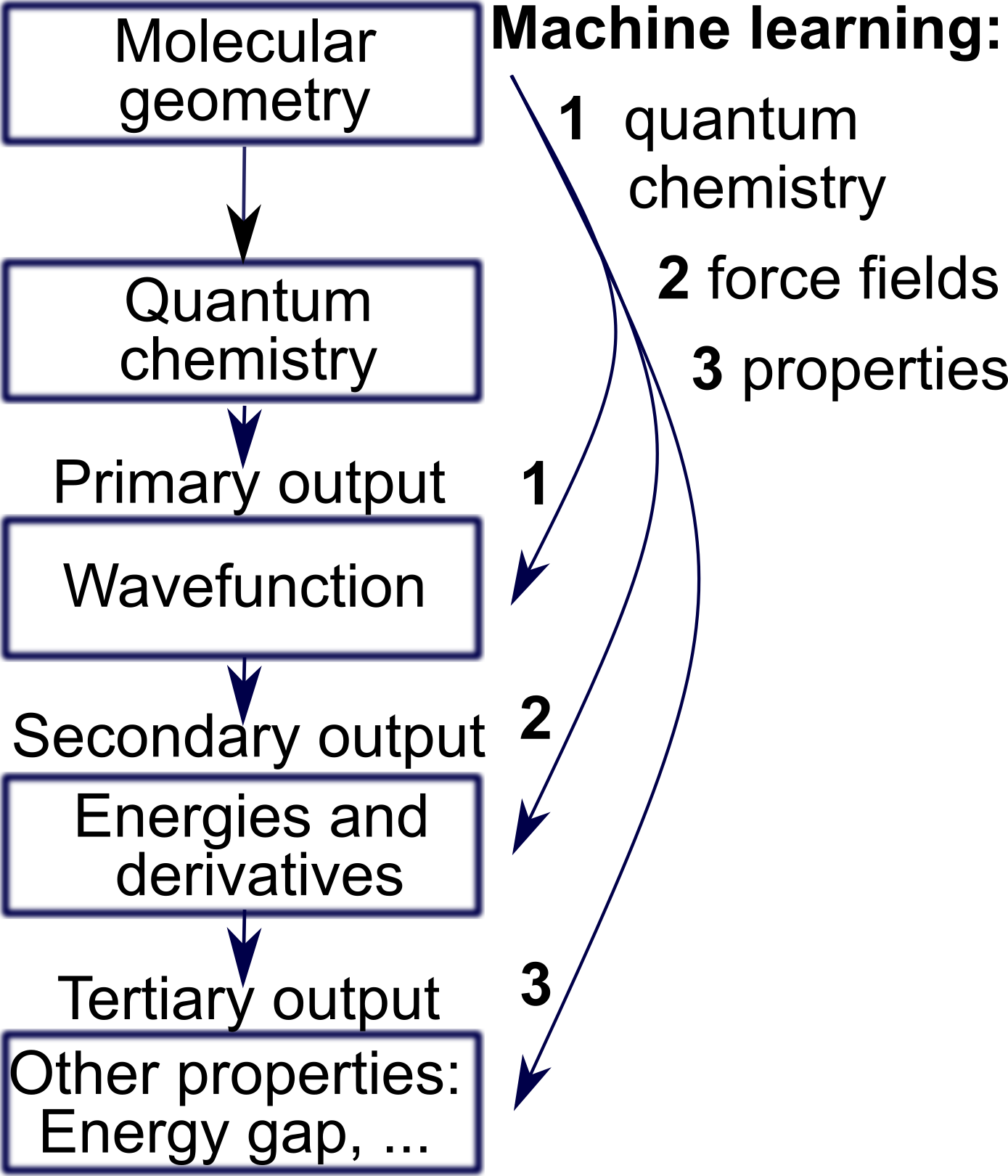}
    \caption{A broad classification of how to use machine learning models to replace different parts of quantum chemistry~\cite{Chandrasekaran2019nCM}. Simulations can be enhanced by providing (1) a wavefunction from a machine learning model, (2) a force field by fitting energies and forces or (3) other properties, such as energy gaps or reaction rates by learning the final output of a dynamics simulation directly.machine learning model and   overview of the parts of a quantum chemical simulation.}
    \label{fig:1}
\end{figure}

Technically, the nuclear part and the electronic part of the calculations can be separated to a large extent. First, the electronic problem is solved leading to potential energies for the nuclei. Afterwards, the nuclei move on these potentials classically or quantum chemically~\cite{Doltsinis2006NIC,Koeppel84ACP,Worth2004ARPC,Richter2011JCTC,Mai2018WCMS,Ibele2019MP,Mai2020ACIE}. These two subsequent steps can be carried out in every time step (on-the-fly), if classical trajectories are employed. Alternatively, the two steps are separated as much as possible by precomputing the potential energy surfaces (PESs) and then using these precomputed PESs in the subsequent nuclear dynamics. Experimental observables and macroscopic properties can be obtained in follow-up computations or analysis runs.
Machine learning (ML) can accelerate the overall simulation process on different levels and at several points.
A broad classification of how to use ML models to replace different parts of quantum chemistry to make simulations more efficient is given in Fig.~\ref{fig:1}~\cite{Chandrasekaran2019nCM}.
{}
The probably most fundamental way is to use ML to solve the Schr\"{o}dinger equation. This has been done for the ground state by representing the molecular wavefunction on a grid, in a molecular orbital basis or in a Monte-Carlo approach~\cite{Carleo2017S,Saito2017JPCJ,Nomura2017PRB,Han2018arXiv,Townsend2019JPCL,Schuett2019NC,Pfau2019arXiv,Hermann2019arXiv,Gastegger2020arXiv} and has recently also been applied for the excited states of a one-dimensional model~\cite{Choo2018PRL}. ML can also be used to reconstruct the wavefunction from near-field spectra~\cite{Zheng2019PRL} or to bypass the Kohn-Sham equation in density functional theory (DFT)~\cite{Brockherde2017NC}. The external potential, functional, electronic density or local density of states can be learned~\cite{Hegde2017SR,Brockherde2017NC,Gastegger2019MC,Nelson2019PRB,Chandrasekaran2019nCM,Cheng2019JCP,Lei2019PRM,Zhou2019JPCL,Kolb2017SR}. Very recently, Ceriotti and co-workers further introduced a smooth atomic density by defining an abstract chemical environment~\cite{Willat2019JCP}. 

By having access to the molecular wavefunction or the electron density, the secondary output, which are energies and forces for the ground state and additionally couplings for the excited states, can be derived efficiently with ML. The coefficients of the ML wavefunctions or the density can further be used as an input for quantum chemical simulations, reducing the number of SCF iterations substantially~\cite{Schuett2019NC}.

Instead of learning the quantum chemistry of systems, the so-called "secondary outputs"~\cite{Chandrasekaran2019nCM} can also be mapped directly to a molecular structure, giving rise to so-called ML force fields. By training an ML model on {\emph ab initio} data, the accuracy of quantum chemistry can be combined with the efficiency of conventional force fields for molecular dynamics (MD) simulations in the ground state~\cite{Hobday1999MSMSE,Bartok2010PRL,Rupp2012PRL,Li2015PRL,vonLilienfeld2015IJQC,Gastegger2015JCTC,Rupp2015JPCL,Behler2016JCP,Urban2016CMS,Gastegger2016JCP,Artrith2017PRB,Gastegger2017CS,Deringer2017PRB,Botu2017JPCC,Glielmo2017PRB,Smith2017CS,Fujikake2018JCP,Behler2017ACIE,Zong2018npjCM,Wood2018JCP,Chen2018JCTC,Bartok2018PRX,Chmiela2018NC,Imbalzano2018JCP,Zhang2018NIPS,Zhang2018PRL,Chan2019JPCC,Faber2018JCP,Wang2019JCTC,Gerrits2019JPCL,Chmiela2019CPC,Carleo2019RMP,Krems2019PCCP,Deringer2019AM,Ward2019MRSC,Noe2020ARPC}. For the excited states, only a couple of studies are available~\cite{Behler2008PRB,Carbogno2010PRB,Hu2018JPCL,Dral2018JPCL,Chen2018JPCL,Williams2018JCP,Xie2018JCP,Guan2019PCCP,Westermayr2019CS,Guan2020JCTC,Krems2019PCCP,Richings2018JCP,Alborzpour2016JCP,Richings2019JCTC,Richings2019JCTC,Polyak2019JCP,Guan2019JCP,Wang2019JPCA,Guan2020JPCL,Richings2018JCP,Richings2017CPL,Richings2017CPL,Netzloff2006JCP,Bettens1999JCP}. Nevertheless, the first NAMD simulation with ML dates back to the year 2008, where the scattering of O$_2$ on Al(III) was studied in a mixed quantum-classical approach considering singlet-triplet transitions~\cite{Carbogno2008PRL,Carbogno2010PRB}. 

Having access to the excited state energies, "tertiary properties", such as UV spectra~\cite{Ghosh2019AS}, band gaps~\cite{Pilania2017CMS,Schuett2017NC,Zhou2018JPCL}, HOMO-LUMO gaps or vertical excitation energies~\cite{Pereira2017JCIM,Isayev2017NC,Pronobis2018EPJB,Stuke2019JCP} of molecules can be derived. Again, this tertiary output can also be fit in a direct fashion, which has been done for instance for a light-harvesting system by learning the excitation energy transfer properties~\cite{Haese2016CS} or the output of NAMD simulations to find out about the relations of molecular structures and dynamic properties~\cite{Haese2019CS}.
Moreover, ML has been successfully applied for the inverse design of molecules and materials featuring specific properties, such as defined HOMO-LUMO gaps or catalytic activities. Examples range from the inverse design of photonic materials, to (photo-)catalysts, solar cells or (photo-active) drugs, to name only a few applications~\cite{Boyle2011JPCC,Teunissen2017JCTC,Liu2018ACSP,Elton2019MSDE,Deringer2019AM,Sanchez-Lengeling2018s,Goldsmith2018AJ,Davies2018FD,Lilienfeld2019arXiv,Freeze2019CR,Lee2020OE}.

Despite the opportunities of ML for the development of groundbreaking new methodologies, current techniques are often limited to certain molecules or specific problems. Methods exist, that extrapolate throughout chemical compound space, see e.g. Refs.~\cite{Montavon2013NJP,Hansen2015JPCL,Huang2016JCP,Christensen2019JCP,Christensen2020JCP,Schuett2018JCP,Schuett2019NC,Tawfik2019ATS,Christensen2019CHIMIA,Fias2019JPCL,Ghosh2019AS,Rudorff2020PCCP}, but usually models fail to go beyond energies and related properties, such as forces, atomization or excitation energies. Further, it is challenging to predict compounds consisting of atom types strongly different from those inside of the training set.
Especially the fitting of the excited-state PESs poses another obstacle, let alone the transferability of excited-state PESs: Not only are ML models restricted to certain molecules or materials~\cite{Behler2008PRB,Carbogno2010PRB,Hu2018JPCL,Dral2018JPCL,Chen2018JPCL,Williams2018JCP,Xie2018JCP,Guan2019PCCP,Westermayr2019CS,Guan2020JCTC,Krems2019PCCP,Richings2018JCP,Alborzpour2016JCP,Richings2019JCTC,Richings2019JCTC,Polyak2019JCP,Guan2019JCP,Wang2019JPCA,Guan2020JCTC,Richings2018JCP,Richings2017CPL,Richings2017CPL,Netzloff2006JCP,Bettens1999JCP}, more often the different energetic states are fit independently from each other with separately trained ML models. As it is clear that the PESs of molecules are not independent of each other, it might also be unsurprising that learning them simultaneously is advantageous for various applications, such as NAMD or spectra predictions. Only a few studies exist that include more energetic states in one ML model and even less treat related properties, such as the vectorial dipole moments or couplings between different PESs in one ML model~\cite{Gastegger2017CS,Ghosh2019AS,Christensen2019JCP,Westermayr2019CS,Westermayr2020MLST,Zubatyuk2019SA}.

However, in our view, the "holy grail" of ML for photochemistry is an ML model that provides all relevant energetic states, forces and properties at once, using derivatives where possible, rather than learning the properties independently. At the very best, this model should be transferable throughout chemical compound space~\cite{vonLilienfeld2018ACIE} and could be used for molecules of any composition. Given the fact that ML models for the electron wavefunctions of different excited states (or related models within the DFT framework) do not exist for polyatomic systems, this dream has not yet come alive. Hence, we will focus this perspective on ML models that learn the secondary outputs, i.e., excited state PESs, corresponding forces, and nonadiabatic and spin-orbit couplings (NACs and SOCs, respectively) between them. We note that our discussion holds for different spin-multiplicities, although most studies focus on singlet states only. We try to address the recent achievements in the fields of photochemistry using ML and discuss the current challenges and future perspectives to get a step further to a transferable ML model for excited states that treats all properties on the same footing.

We start by discussing the generation of a training set for the treatment of excited state PESs, corresponding forces and couplings and focus on their use in NAMD simulations. Especially, we aim to clarify the differences between excited-state and ground-state properties. We therefore describe the NACs and SOCs that couple different electronic states and highlight their importance for NAMD simulations. Subsequently, state-of-the-art ML models for excited-state PESs are considered along with the challenge of modelling a manifold of energetic states.

\section{Generating a training set for excited states}
The basis of any successful ML model is a comprehensive and accurate training set that contains the molecular geometries in combination with the corresponding properties that need to be predicted. For the application of ML within NAMD simulations, the training set should contain a molecular geometry and the energies of all energetic states, corresponding forces, and couplings between the states. It is computed with the quantum chemistry method, whose accuracy one wants to obtain. The choice of the quantum chemistry method is a problem on its own and often requires expert knowledge~\cite{Gonzalez2020,Park2020CR}. Simply said, ML can be seen as efficient interpolation between data points with the accuracy of the reference method. 

Before we go into detail on how to efficiently create a training set for excited states, we will first discuss the differences to ground state potentials and properties that need to be considered. A major drawback is the fact that a manifold of excited states and thus also the properties between them have to be accounted for. These are NACs between states of same spin multiplicity and SOCs between states of different spin multiplicity as well as transition dipole moments.
The fitting of such properties is problematic due to the arbitrary phase of the wavefunction \cite{Westermayr2019CS,Westermayr2020MLST}. Therefore, an additional pre-processing might be necessary. Either a diabatization~\cite{Williams2018JCP,Xie2018JCP,Guan2019PCCP,Guan2020JCTC,Krems2019PCCP,Richings2018JCP,Alborzpour2016JCP,Richings2019JCTC,Richings2019JCTC,Polyak2019JCP,Guan2019JCP,Wang2019JPCA,Guan2020JPCL,Richings2018JCP,Richings2017CPL,Richings2017CPL}, a so-called phase correction~\cite{Westermayr2019CS}, or a special learning algorithm~\cite{Westermayr2020JPCL} renders data learnable. The latter two are described in the following while further details on the former are given in section~\ref{diabatic}.

\subsection{Making excited-state data learnable}
Compared to energies and forces, NACs, SOCs as well as transition dipole moments result from the wavefunctions of two different electronic states. Due to the non-unique definition of the wavefunction itself, i.e., the fact that multiplication of the electronic wavefunction with a phase factor still gives a valid eigenfunction of the electronic Hamiltonian, leads to an arbitrary phase, which is initiated randomly in a quantum chemical calculation. Consequently, also the sign of the couplings, $C_{ij}$, can be positive or negative. Here, $i$ and $j$ denote the indices of the involved states. The resulting inconsistencies in the coupling hypersurfaces make it challenging to find a good relation between an ML model, which is per definition a smooth function~\cite{Goodfellow2016}, and those discontinuous raw outputs. 

\begin{figure}[ht]
\centering
    \includegraphics[scale=0.4]{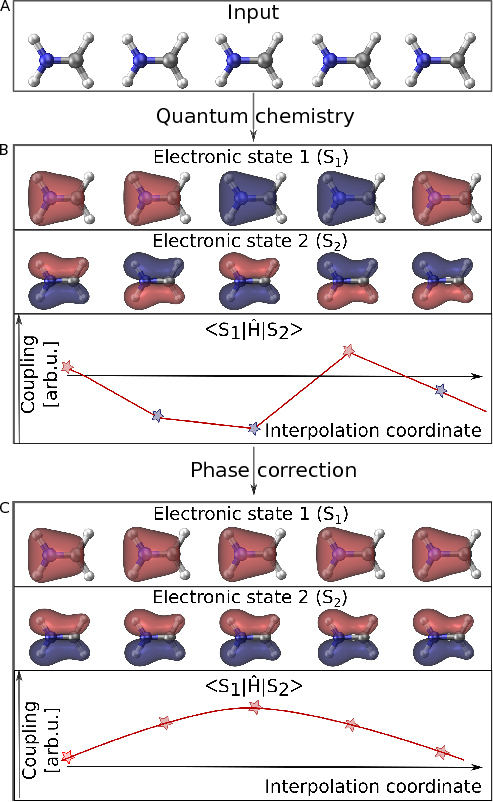}
    \caption{A set of quantum chemical calculations of the methylenimmonium cation, CH$_2$NH$_2^+$, along the C-N bond. The molecular conformations, which are the input of a quantum chemical calculation, are given in panel A, the orbitals of the computed two electronic states, the $S_1$ and $S_2$ state, as well as the corresponding off-diagonal matrix elements between those two states, $\langle S_1 \mid \hat{H}\mid S_2\rangle$, are given in panel B. As can be seen, the sign of those values arbitrarily switches. Those sign jumps can be removed by applying a phase correction algorithm. Results are given for those elements in panel C. Reproduced from Ref.~\cite{Westermayr2019CS} under CC-BY, https://creativecommons.org/licenses/by/3.0/.}
    \label{fig:phase}
\end{figure}

This problem can be illustrated with molecular orbitals of the methylenimmonium cation (Fig.~\ref{fig:phase} reproduced from Ref.~\cite{Westermayr2019CS}). Panel A shows the molecular geometries, which are given as an input to a quantum chemical program. Two molecular orbitals are shown as placeholders for the wavefunctions of two electronic states, the $S_1$ and $S_2$ states. The color of the orbitals can be either blue or red and changes arbitrarily throughout the reaction coordinate. In the same way, also the overall wavefunction (which is difficult to plot) for the respective state changes its phase arbitrarily. As a consequence, also the sign of the couplings, where the product of the two wavefunctions' signs enters, may change randomly (see panel B).

In order to allow for a meaningful ML description of elements resulting from two different electronic states, a data pre-processing is helpful. 
The former process is termed phase correction~\cite{Akimov2018JPCL,Westermayr2019CS} and is practicable to remove almost all such inconsistencies in the configurational space of the training set. This phase correction makes the use of conventional training algorithms possible. 

To carry out phase correction, a wavefunction overlap computation~\cite{Plasser2016JCTC}, $\boldsymbol{S}=\langle \Psi_\alpha \mid \Psi_\beta\rangle$, has to be carried out between the wavefunction $\Psi_\beta$ at every geometry $\beta$ inside the training set and the wavefunction $\Psi_\alpha$ at a reference geometry $\alpha$. The phase thus has to be tracked from a pre-defined reference geometry. It often happens that two geometries are dissimilar to each other, so that interpolation between them is necessary, making this process generally more expensive. So-called intruder states give rise to additional problems, since they are so high in energy at the reference geometry that they usually would not be included in the initial calculation. However, they enter the lower energy region at another geometry visited during the NAMD simulations and thus need to be considered in the current calculation step. Hence, they should have been included from the beginning for the phase correction algorithm to work. As a solution to this problem, many electronic states need to be computed from the start. In some cases, where many energetic states lie close to each other and where the photochemistry is complex, phase correction might even be infeasible. The problem of intruder states was also identified by Robertsson et al. and is well explained for a diabatization procedure in Ref.~\cite{Robertson2019JCC}. For a more detailed discussion on phase correction, the reader is referred to Ref.s~\cite{Westermayr2019CS,Mai2015IJQC,Plasser2016JCTC,Akimov2018JPCL}.
Nevertheless, as given in panel C of Fig.~\ref{fig:phase}, smooth curves are obtained if phase correction is carried out correctly and these phase-corrected properties can be learned with conventional ML models. 

Similarly, a small set of data can be corrected manually and afterwards a cluster growing algorithm can be applied~\cite{Shu2019JCP,Wang2019JPCA}. This algorithm uses Gaussian process regression to continuously add data points to the initially phase-corrected data set. This approach has been employed recently to obtain diabatic transition dipole moments~\cite{Guan2020JCTC}. However, in systems containing many degrees of freedom and many electronic states, a manual correction of the sign of couplings is tedious and the approach has only been applied to small systems, yet~\cite{Guan2020JPCL,Guan2020JCTC}.

In contrast to the phase-correction procedures described above, an ML-based internal phase-correction during training renders the learning of raw quantum chemical data possible and does not require any pre-processing. However, it requires a modification of the training process itself~\cite{Westermayr2020JPCL}. 
In a recent study, we applied such a phase-free training using the deep continuous-filter convolutional-layer NN SchNet~\cite{Schuett2018JCP,Schuett2019JCTC} that we adapted for the treatment of excited states. In contrast to conventional algorithms, where the hyperparameters of the network are optimized to minimize the L$_1$ or L$_2$ loss function, here a phase-less loss function is applied. The latter allows the ML model to possess a different phase (or sign) for the learned property than the reference data. Since ML models intrinsically yield smooth curves, the algorithm will then automatically choose a phase for every data point such that smooth coupling curves are produced. This freedom of choice is achieved by calculating the errors between the ML value and all possible phase variations of the reference value and using only the smallest of these errors. The possibilities for phase conventions scales with $2^{N_S-1}$, where $N_S$ is the number of considered states. Since the error is computed more often than in conventional ML training, the phase-less loss training becomes more expensive, when including more electronic states.
Mathematically, instead of computing the mean squared error, $\varepsilon_{L_2}$, between reference couplings, $C_{ij}^{QC}$, and predicted couplings, $C_{ij}^{ML}$, 
\begin{equation}
    \varepsilon_{L_2} = \mid \mid C_{ij}^{QC} - C_{ij}^{ML} \mid\mid^2,
\end{equation}
the phase-free error, $\varepsilon_{ph}$, is computed as the minimum of $2^{N_S-1}$ computed errors:
\begin{equation}
     \varepsilon_{ph} = \mid\mid C^{QC}_{ij} \cdot p_i^{k} \cdot p_j^{k} - C^{ML}_{ij} \mid\mid  ~\text{with}~ 0 \leq k \leq 2^{N_S-1}.
\end{equation}
$p_i^k$ and $p_j^k$ are phase factors, giving rise to the sign of state $i$ and $j$ resulting in a phase vector with index $k$. This adaption of the loss function can remove the influence of any phase during the training process, making the use of raw data possible. A variation of this approach with reduced cost is possible if only one property, i.e. NACs or SOCs, are trained for NAMD simulations. A detailed discussion can be found in Ref.~\cite{Westermayr2020JPCL}. 

It is worth mentioning that, besides the arbitrary phase of the wavefunction, also the Berry phase (or geometric phase)~\cite{Ryabinkin2017ACR} exists. Effects due to the Berry phase can not be accounted for with phase correction. Nevertheless, most often in mixed quantum classical NAMD simulations, the Berry phase can be neglected. As a drawback, some effects, such as interference of nuclear wavefunctions might not be described correctly with such methods, and thus prevents the application of the corresponding ML properties if those effects are important. In some other dynamics methods and reactions, the Berry phase plays a crucial role and can lead to path-dependent transition probabilities close to conical intersections. This effect is important in quantum dynamics simulations, where problems can be circumvented by using diabatic potentials, which will be described in section~\ref{diabatic}.

\subsection{Choosing the right reference method}

While many ML potentials for ground-state MD simulations are based on DFT training sets, see e.g. Refs.~\cite{Behler2008PRL,Li2015PRL,Gastegger2015JCTC,Gastegger2018JCP,Deringer2017PRB}, the training sets for the excited states are mainly obtained with multi-reference methods. Examples are the complete active space self-consistent-field (CASSCF) method~\cite{Richings2017CPL,Hu2018JPCL,Chen2018JPCL,Dral2018JPCL,Richings2018JCP,Richings2019JCTC,Haese2019CS,Westermayr2020JPCL} or multi-reference configuration interaction (MR-CI) schemes~\cite{Koch2014JCP,He2016SR,Guan2017JCP,Williams2018JCP,Wang2018SR,Yuan2018PCCP,Yin2019PCCP,Guan2019PCCP,Westermayr2019CS,Westermayr2020MLST,Schwilk2020arXiv,Guan2020JPCL,Guan2020JCTC,Westermayr2020JPCL}. The advantage of multi-reference methods compared to single reference methods is that photo-dissociation, which is likely to occur in many molecules after their excitation by light, can be treated accurately. In contrast, single-reference methods fail to do so in many cases. However, multi-reference methods are seriously limited by their computational costs~\cite{Nelson2020CR,Lischka2018CR}, calling for an efficient and meaningful training set generation. Therefore, the training set should be as small as possible, but should cover the relevant conformational space of a molecule that is required for accurate NAMD simulations~\cite{Dral2020JPCL}. Accordingly, many recent training sets for MD simulations are built by a so-called "active-learning"~\cite{Botu2015IJQC,Li2015PRL} or iterative/adaptive sampling scheme~\cite{Behler2015IJQC,Gastegger2017CS} that will be described in the following and can be adapted for excited states~\cite{Westermayr2019CS}. From our point of view, it is most favorable to start by computing a small initial training set and to expand it via such an adaptive sampling scheme~\cite{Li2015PRL,Gastegger2017CS,Westermayr2019CS,Behler2015IJQC,Botu2015IJQC,Smith2018JCP}. 

\subsection{Initial training set}
If only static calculations are targeted, data bases can be generated efficiently by starting from already existing data sets. As an example, Schwilk et al.~\cite{Schwilk2020arXiv} constructed a large data set of 13k carbene structures by randomly choosing 4,000 geometries from the QM9~\cite{Ramakrishnan2014SD} data set (consisting of 130k organic molecular structures). Hydrogen-atoms were abstracted and singlet and triplet states were optimized. The MR-CI method was subsequently used to compute the energies of the singlet and triplet state and a data set of 13,000 different carbene structures, called QMspin, was obtained, opening avenues to investigate important intermediate geometries critical for organic reaction networks.

As a starting point for all following training-set generation schemes aiming to investigate the temporal evolution of a system, the equilibrium geometry of a molecule can be computed and taken as a reference. The initial training set can then be built up by sampling conformations close to this molecular configuration. In general, every sampling method is possible. Since the normal modes of a molecule are generally important for dynamics, scans along these coordinates can be used to sample different conformations. In cases of small molecules with few degrees of freedom, this process might be a good starting guess for an initial training set~\cite{Westermayr2019CS}. It also makes sense to optimize critical points like excited-state minima, conical intersections and state crossings and to include data along such optimization runs into the training set. The same is advisable for larger systems, but in addition, some other approaches like Wigner sampling~\cite{Wigner1932PR} or sampling via MD simulations~\cite{Bruccoleri1990B,Maximova2016PLOSCB} can be considered. To name a few approaches, umbrella sampling~\cite{Kaestner2011WIRCMS}, trajectory-guided sampling~\cite{Tao2019TCA}, enhanced sampling~\cite{Yang2019JCP} or metadynamics~\cite{Herr2018JCP}, using a cheap electronic structure method like the semi-empirical tight-binding based quantum chemistry method GFN2-xTB~\cite{Grimme2019JCTC}), can be employed.

Further, if literature or chemical intuition indicate that certain reactions, like dissociation, take place after photo-excitation, it is also favorable to include those reaction coordinates right from the beginning. The initial training set can easily comprise on the order of 1,000 data points, which might seem like a lot but is reasonable given the large number of data points in commonly used training sets~\cite{Chen2018JPCL,Hu2018JPCL,Dral2018JPCL,Westermayr2020MLST}. The quality of the initial ML potentials can be assessed by carrying out short scans along different reaction coordinates, such as combinations of normal modes. As soon as the initial training set is large enough, the training set expansion via an adaptive sampling scheme can be started.

\subsection{Adaptive sampling for excited-states}

ML models fail to predict regions with scarce training data, i.e., their extrapolation capabilities are faint~\cite{Rupp2015IJQC}. Since such regions are likely visited during a dynamics simulation, the initial training set then needs to be expanded. A quality control is needed to detect whether unknown conformational regions of the molecule are visited, such that the corresponding structures afterwards can be added to the training set.

This concept was introduced already in 1992 as query by committee~\cite{Seung1992} and has been used in chemistry in the so-called GROW algorithm of Collins and coworkers~\cite{Bettens1999JCP,Netzloff2006JCP} as well as in the iterative sampling of Behler~\cite{Behler2015IJQC}. The latter is nowadays well adapted for the ground state~\cite{Botu2015IJQC,Li2015PRL,Gastegger2017CS,Smith2018JCP} and was recently modified for the excited states~\cite{Westermayr2019CS}. The scheme is described in more detail in the following. 

\begin{figure}[ht]
    \centering
    \includegraphics[scale=0.39]{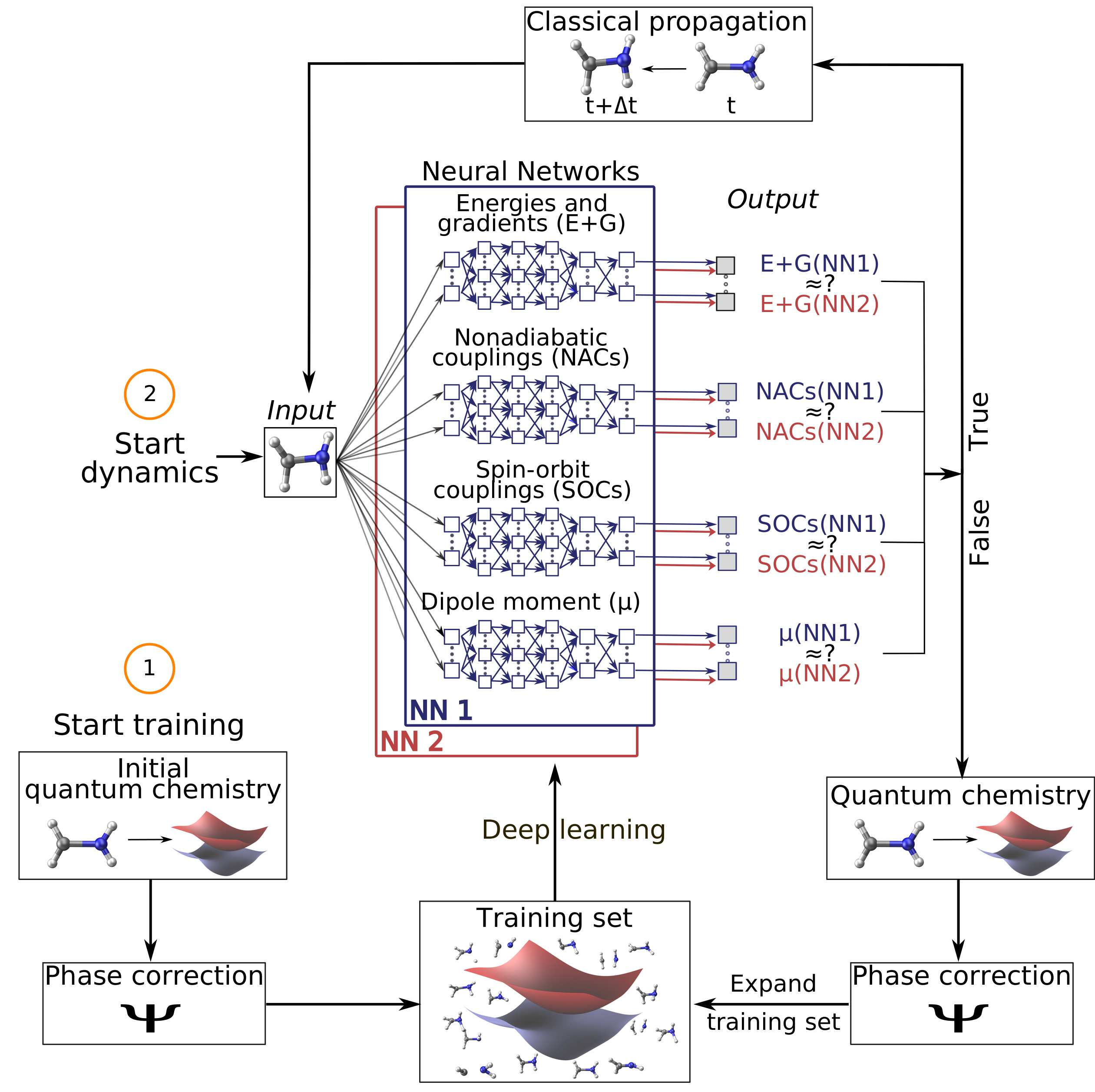}
    \caption{Overview of the adaptive sampling scheme for excited states, reproduced from Ref.~\cite{Westermayr2019CS} under CC-BY. Adaptive sampling is illustrated exemplarily with two NNs for the methylenimmonium cation, CH$_2$NH$_2^+$. As a starting point, ML models are trained on the initial training set and ML-based dynamics are executed. At each time step, the predictions of the two deep NNs (NN1 and NN2) are compared to each other for energies (E) and gradients (G), nonadiabatic couplings (NACs), spin-orbit couplings (SOCs) and dipole moments ($\mu$). If the difference between the ML models overcomes a pre-defined, adaptive threshold, the geometry visited at this time step is re-computed with quantum chemistry, added to the training set after phase correction and the ML models are re-trained. Subsequently, a new dynamics cycle is started and this process is repeated until the ML models are deemed to be converged.}
    \label{fig:adaptivesampling}
\end{figure}

To apply the procedure of adaptive sampling, at least two ML models have to be trained independently, e.g., with slightly different hyperparameters or starting weights. An overview of this procedure with two neural networks (NNs) is given as an example in Fig.~\ref{fig:adaptivesampling}. 
At every time step during an MD simulation, the predictions ${Y}_M^p$ of at least two ML models, $M$, for a property $p$ (e.g., a potential energy) are compared to each other.
To this end, the standard deviation of these predictions with respect to the mean of each property, $\overline{Y}^P$, is computed according to 
\begin{equation}
Y_\sigma^p = \sqrt{\frac{1}{M-1}\sum_{m=1}^J ( Y^p_m-\overline{Y}^p)^2}.
\end{equation}
This standard deviation is compared to a pre-defined threshold, $\varepsilon^p$, for each trained property.
If the standard deviation stays below the threshold, the mean of each property, $\overline{Y}^P$, is forwarded to the MD program to propagate the nuclei.
If the threshold is exceeded, the ML prediction is assumed to stem from an undersampled or unknown region of the PESs and is deemed untrustworthy. This conformation has to be included into the training set to guarantee accurate ML PESs. Thus, a quantum chemical reference computation is carried out, the data point is added to the training set and the ML models are re-trained to execute ML-NAMD simulations on longer time scales. It is sensible to choose a large threshold, $\varepsilon^p$, in the beginning and adaptively make it smaller as the robustness of the ML models increases, giving rise to the name adaptive sampling~\cite{Gastegger2017CS}.

Adaptive sampling for excited states differs from adaptive sampling in the ground state in the number of properties that are considered. As illustrated in Fig.~\ref{fig:adaptivesampling}, not only the energies must be accurately predicted, but also the couplings and, if necessary, dipole moments.
Since more states are considered, an average standard deviation is taken as the mean of the standard deviations of each state in case of energies and gradients,
    \begin{equation}
Y_\sigma^p = \frac{1}{N_S}\sum_i^{N_S}\left(\sqrt{\frac{1}{M-1}\sum_{m=1}^M ( Y^{p_i}_{m}-\overline{Y}^{p_i})^2}\right),
\end{equation}
and as the mean of the standard deviations of each pair of states for couplings or dipole moments,
   \begin{equation}
Y_\sigma^p = \frac{1}{2N_S^2}\sum_{i}^{N_S}\sum_j^{N_S}\left(\sqrt{\frac{1}{M-1}\sum_{m=1}^M ( Y^{p_{ij}}_{m}-\overline{Y}^{p_{ij}})^2}\right).
\end{equation}
A separate threshold is set for each of these averaged quantities. If any of the quantities is predicted inaccurately, the data point is recomputed with quantum chemistry, phase corrected and added to the training set.

In order to make this process more efficient, not only one MD simulation, but an ensemble of trajectories can be computed. The ML models are only retrained after each of the independent trajectories has reached an untrustworthy region of the PES and after each of the reference calculations has been finished and included in the training set. This makes the parallelization of many trajectories possible~\cite{Behler2015IJQC,Gastegger2017CS,Westermayr2019CS,Gastegger2018arXiv}. 

The adaptive sampling scheme should be carried out until the relevant conformational space for photodynamics is sampled sufficiently. However, using more than one ML model for production runs is still favorable. One of us and coworkers observed that the error of predictions decreases with the number of ML models used~\cite{Gastegger2017CS,Gastegger2018arXiv}. We have seen the same trend in a recent study for NAMD simulations. With the adaptive sampling scheme for excited states, we generated a training set of 4,000 data points of the methylenimmonium cation, CH$_2$NH$_2^+$, to carry out long time-scale NAMD simulations with NNs~\cite{Westermayr2019CS}.

\subsection{Additional sampling techniques for excited states}
Further training sets for NAMD simulations were generated for one-dimensional systems as well as polyatomic molecules. For example, Chen et al. have computed 90,000 data points via Born-Oppenheimer MD simulations and NAMD simulations, where emphasis was placed on the inclusion of geometries after a transition from one state to another took place~\cite{Chen2018JPCL}. Deep NNs were trained on energies and gradients of this training set to accurately reproduce NAMD simulations.

Hu et al.~\cite{Hu2018JPCL} used a very similar approach and obtained around 200,000 data points of 6-aminopyrimidine from Born-Oppenheimer MD simulations. They further carried out NAMD simulations with surface hopping~\cite{Tully1990JCP,Tully1991IJQC}, where transitions from one state to another were allowed via so-called hops. The geometries that were visited shortly before a hop took place were used as a starting guess to optimize conical intersections, i.e. critical points of the PESs, where two states become degenerate. Those data points were included to comprehensively sample the regions around a conical intersection. However, the ML models were not accurate enough for NAMD simulations solely based on ML potentials and the authors had to resort to quantum chemistry calculations in critical regions of the PESs.

Dral et al.~\cite{Dral2018JPCL} generated a training set for a one-dimensional two-state spin-boson model consisting of 10,000 data points with a grid-based method. The training data selection was then based on the structure, rather than on the energy of the molecules. For each data point, a molecular descriptor was computed and the distances of the descriptors were compared. Data points for the training set were chosen to sample the relevant space sufficiently~\cite{Dral2017JCP,Dral2018JPCL}. Compared to random sampling, this method allowed a reduction of training set sizes up to 90 \%, which was shown for static calculations of the methyl chloride molecule~\cite{Dral2017JCP,Sobol2011W}. A similar structure-based sampling scheme was proposed by Ceriotti et al.~\cite{Ceriotti2013JCTC}.

Additionally, a maximum and minimum value can be computed for each representation of a molecule inside the training set. Every new structure that is obtained throughout an MD run can be compared to those values to get a measure of reliability of ML predictions. If the configuration does not lie within the known region, it can be added to the training set~\cite{Botu2015IJQC,Behler2011PCCP}.
Very recently, an active learning approach has been proposed to construct PESs without the need of running MD trajectories. The difference between two NN potentials was computed and points were iteratively added at the maxima of this difference surface (or, as phrased in the study, at the minima of the negative difference)~\cite{Lin2020JCP}.

It becomes evident from the diversity of approaches and training set sizes that a general guide on how to compute the training set and how large it should be for NAMD simulations can not be given. It is rather a matter of the efficiency that should be achieved and the computational costs that can be justified.
Further, the training set strongly depends on the molecule under investigation. Especially its size, flexibility and the complexity of the light-induced dynamics play an important role. A molecule, whose photodynamics can be described as a two-state problem, such as in a simplified case of ethylene~\cite{Barbatti2005JCP,Hollas2018JCTC}, or a molecule, which is rigid, where dynamics mostly lead towards one reaction channel, possibly requires less data points than molecules that exhibit several different reaction channels after photo-excitation. 

\section{Machine learning nonadiabatic molecular dynamics simulations (ML NAMD)}

\subsection{Beyond Born-Oppenheimer dynamics}
With an accurate training set for excited states at hand, NAMD simulations can be enhanced with ML models in order to enable the dynamics on time scales otherwise unfeasible. 
The most accurate way to study the dynamics of a molecule would be the full quantum mechanical treatment, which is, however, expensive and limited to a few atoms, even if ML PESs are applied~\cite{Koeppel84ACP,Koeppel2001JCP,Bowman2008MP,Meyer2009,Alborzpour2016JCP,Richings2017CPL,Liu2017SR,Richings2017CPL,Richings2018JCP,Williams2018JCP,Xie2018JCP,Guan2019PCCP,Richings2019JCTC,Polyak2019JCP,Guan2019JCP,Wang2019JPCA,Guan2020JCTC}. A mixed quantum classical treatment is thus often preferred, where the motion of the nuclei are treated classically on one of the ML PESs. The mixed quantum classical MD simulation can then be interpreted as a mixed MLMD simulation~\cite{Westermayr2020JPCL}. The Born-Oppenheimer approximation allows to separate the nuclear from the electronic degrees of freedom. However, this approximation is not valid in the vicinity of avoided state crossings of PESs (or conical intersections, as mentioned before), which play an important role in excited-state dynamics.

In these critical regions of the PESs, ultrafast rearrangement of the motions of the electrons and the nuclei takes place due to strong couplings.  
As already mentioned, the relevant coupling elements are NACs and SOCs. The NACs (denoted as $C^{\text{NAC}}$) are vectorial properties and can be computed as~\cite{Baer2002PR,Lischka2004JCP,Doltsinis2006NIC} 
\begin{equation}
\begin{array}{lr}
    \label{eq:nac5}
        C^{\text{NAC}}_{ij} \approx \langle  \Psi_i \mid \frac{\partial}{\partial \mathbf{R}}\Psi_j \rangle =\\
        \frac{1}{E_i-E_j}{\langle \Psi_i \mid \frac{\partial H_{el}}{\partial \mathbf{R}}\mid \Psi_j \rangle }   ~~~\text{for}~  i \neq j,
\end{array}
\end{equation}
neglecting the second order derivatives. Thus, in the vicinity of a conical intersection, the couplings become very large, whereas they are almost vanishing elsewhere. The singularities that arise when two states are degenerate do not only pose an obstacle to quantum chemistry, but consequently also to PESs fitted with ML~\cite{Dral2018JPCL,Hu2018JPCL,Chen2018JPCL,Westermayr2019CS}. NACs are nevertheless important properties to determine the direction and probability of internal conversion -- a transition from one state to another, where the spin multiplicity does not change~\cite{Doltsinis2006NIC,Richter2011JCTC,Mai2018WCMS,Ha2018JPCL,Mai2020ACIE}. 
In contrast, the SOCs (denoted as $C^{\text{SOC}}$) are complex-valued properties that determine the rate of intersystem crossing, i.e., the transitions from one state to another, where spin multiplicity does change. In standard quantum chemistry programs, SOCs are given as the off-diagonal elements of the Hamiltonian matrix~\cite{Mai2018WCMS,Granucci2012JCP,Mai2020ACIE}: 
\begin{equation}
C^{\text{SOC}}_{ij}=\langle \Psi_i \mid \hat{H}^{SOC}\mid \Psi_j \rangle .
\end{equation}

\subsection{Fitting diabatic potentials}\label{diabatic}
The numerical difficulties that arise due to discontinuous PESs and singularities of couplings at conical intersections can be circumvented by the use of diabatic potentials instead of adiabatic ones~\cite{Tannor2006,Xie2018JCP,Jasper2004,Yarkony2005JCP,Zhu2016JCP,Wittenbrink2016JCP}. In the diabatic basis, the coupling elements are smooth properties and the arbitrary phase of the wavefunction does not have an impact. This favors the use of diabiatic PESs. Since the output of a quantum chemistry program is generally given in the adiabatic basis, a quasi-diabatization procedure is necessary. Strictly speaking, a diabatization procedure is not possible because e.g. an infinite number of states is needed for an accurate representation. If using a finite number of states, the term quasi-diabatic is employed. For simplicity, we still use the notation of diabatic potentials for quasi-diabatic potentials. Those have been generated with different methods~\cite{Domcke2004} and for small molecules up to date. Examples are the propagation diabatization~\cite{Richings2015JPCA}, diabatization by localization~\cite{Accomasso2019CPC} or by ansatz~\cite{Lenzen2017JCP,Williams2018JCP}, diabatization based on couplings or other properties~\cite{Subotnik2008JCP,Hoyer2016JCP,Wittenbrink2013JPCA,Varga2018PCCP}, configuration uniformity~\cite{Nakamura2002JCP}, block-diagonalization~\cite{Venghaus2016JCP}, CI vectors~\cite{Robertson2019JCC} or (partly) on ML~\cite{Li2013JCP,Jiang2013JCP,Jiang2014JCP,Jiang2016IRPC,Lenzen2017JCP,Williams2018JCP,Xie2018JCP}.

Since several years, (modified) Shepard interpolation is used to fit diabatic potentials~\cite{Ischtwan1994JCP,Bettens1999JCP,Evenhuis2004JCP,Evenhuis2011JCP,Mukherjee2013JPCA} and also least squares fitting was applied to study the photo dissociation of molecules, such as NH$_3$ and phenol~\cite{Zhu2012JCP,Zhu2014JCP}.
In a series, Guo, Yarkony and co-workers developed invariant polynomial NNs~\cite{Li2013JCP,Jiang2013JCP,Jiang2014JCP,Jiang2016IRPC,Xie2018JCP,Guan2019JCP} to address the excited-state dynamics of NH$_3$ and H$_2$O by fitting diabatic potential energy matrix elements. Absorption spectra as well as branching ratios could be obtained with high accuracy. 
The same authors further fit the diabatic 1,2$^1$A dipole moment surfaces of NH$_3$, which can only be fitted accurately if the topography of the PESs is reproduced correctly, validating the previously fitted diabatic potentials~\cite{Guan2020JCTC}. 

Habershon, Richings and co-workers used Gaussian process regression (in their notation equal to kernel-ridge regression) to fit diabatic potentials to execute on-the-fly dynamics of the butatrien cation with variational Gaussian wavepackets~\cite{Richings2017CPL}. In another study, they applied an on-the-fly MCTDH scheme (DD-MCTDH) and carried out 4-mode/2-state simulations of pyrazine~\cite{Richings2018JCP}. By improving the ML approach with a systematic tensor decomposition of kernel ridge regression, the study of 12-mode/2-state dynamics of pyrazine was rendered possible. This achievement remains a huge improvement over current MCTDH simulations in terms of accuracy and efficiency~\cite{Richings2018JCTC}.

For the improvement of the diabatization by ansatz procedure, Williams et al.~\cite{Williams2018JCP} applied NNs and enabled the fitting of the electronic low lying states of NO$_3$.
The improvement of the diabatization procedure itself is desirable~\cite{Xie2018JCP,Guan2019PCCP}, since the generation of meaningful diabatic potentials is often a tedious task and restricts their use tremendously. Up to date, no rigorous diabatization procedure exists that allows the diabatization of adiabatic potentials of polyatomic systems by non-experts in this field~\cite{Xie2018JCP,Guan2020JCTC}. Especially challenging for larger and more complex systems is the number of electronic states within a certain energy range that have to be considered for successful diabatization. An increasing computational effort to provide all relevant electronic states is the result, making diabatization further challenging~\cite{Robertson2019JCC}. Often, more extensive approximations~\cite{Williams2018JCP,Gomez2019JPCA,Robertson2019JCC}, e.g. the linear vibronic coupling model~\cite{Koeppel04} are applied. We refer the reader to Ref.s~\cite{Jasper2004,Koeppel04,Yarkony2004,Worth2004ARPC,Plasser2019PCCP} for more details on such approaches. 

Despite the advantages of diabatic potentials, due to the before-mentioned drawbacks and the fact that the direct output of a quantum chemical calculation is given in the adiabatic basis, on-the-fly NAMD in the adiabatic representation is often the method of choice for large polyatomic systems, which will be discussed in the following.

\subsection{Fitting adiabatic potentials}
In order to execute NAMD simulations in the adiabatic basis, approximations have to be introduced to account for nonadiabatic transitions between different PESs. A good trade-off between accuracy and efficiency can be achieved with the surface hopping methodology~\cite{Tully1990JCP,Tully1991IJQC}, which is often applied in ML-based NAMD simulations~\cite{Hu2018JPCL,Dral2018JPCL,Chen2018JPCL,Westermayr2019CS}. 
In surface hopping, the transitions, or so-called hops, between different states, are computed stochastically and a manifold of trajectories needs to be taken into account to analyze different reaction channels and branching ratios~
\cite{Richter2011JCTC,Fabiano2008CP,Mai2020ACIE}
Several algorithms~\cite{Fabiano2008CP,Oloyede2006JCP,Ishida2017IRPC,Mai2018WCMS} are frequently used to compute the hopping probability as well as its direction, with Tully's fewest switching algorithm being among the most popular ones~\cite{Tully1990JCP,Tully1991IJQC}. There, the couplings between adjacent states determine the hopping probability~\cite{Richter2011JCTC}. Other frequently applied algorithms to compute hopping probabilities are the Landau-Zener~\cite{Zener1932,Wittig2005JPCB} and the Zhu-Nakamura approximations~\cite{Zhu1995PRL,Zhu2002JCP,Oloyede2006JCP,Ishida2017IRPC}. Those approximations solely rely on the PESs and omit the computation of wavefunction coefficients and couplings. Other flavors to account for transitions exist, which have, however, not been used in ML based NAMD studies yet. We thus refer the reader to Ref.s~\cite{Tully1990JCP,Zhu1995PRL,Zhu2002JCP,Doltsinis2006NIC,Granucci2007JCP,Fabiano2008CP,Richter2011JCTC,Malhado2014FC,Mai2015IJQC,Wang2016JPCL,Subotnik2016ARPC,Mai2018WCMS} for further information.

\subsection*{NAMD simulations with ML energies and forces}
Based on the fewest switches algorithm, one of the first ML NAMD simulation is carried out by Carbogno et al.~\cite{Carbogno2008PRL}, where the scattering of O$_2$ at Al(III) is studied~\cite{Behler2007JCP,Carbogno2008PRL,Carbogno2010PRB}. A set of 3768 carefully selected  data points~\cite{phdbehler} allowed for interpolation of the PESs with NNs~\cite{Behler2007JCP}. In a first attempt, the authors include a spin-unpolarized singlet PES and a spin-polarized triplet state. Strictly speaking, the output of a quantum chemistry simulation for a singlet and triplet state is spin-diabatic~\cite{Granucci2012JCP,Mai2018WCMS} and NAMD simulations ideally should carry out a diagonalization to obtain the spin-adiabatic PESs~\cite{Granucci2012JCP,Mai2018WCMS}. The authors took advantage of the adiabatic spin-polarized PES~\cite{Behler2005PRL} to compute the absolute value of couplings~\cite{Carbogno2008PRL}. In this way, the transitions between the states could be approximated using surface hopping omitting the computation of wavefunction dynamics. This study was extended by two-state NAMD simulations of different multiple PESs arising from different spin configurations. Findings suggested a high probability of singlet-to-triplet conversion during scattering experiments with a non-zero probability even at low coupling values~\cite{Behler2007JCP,Carbogno2008PRL,Carbogno2010PRB}.

Other studies using the Zhu-Nakamura method~\cite{Zhu1995PRL,Zhu2002JCP,Oloyede2006JCP,Ishida2017IRPC} to account for nonadiabatic transitions are discussed below. This approximation is based solely on energies and neglects the phase of the wavefunction. As a drawback, PESs are always assumed to couple to each other, when they are close in energy. This holds true for many cases, but one must be aware that strongly and weakly coupled PESs can not be distinguished.

Hu et al.~\cite{Hu2018JPCL} for example trained separate kernel ridge regression models to fit three singlet states of 6-aminopyrimidine. For learning, they used 65,316 data points comprising the molecular structures and energies of 6-aminopyrimidine with gradients not fitted, but computed afterwards. The data points were obtained from Born-Oppenheimer simulations, which were further clustered into sub-groups, from which the training points were selected randomly.  As mentioned before, hopping geometries obtained from reference NAMD simulations were taken to find minimum conical intersections and the latter were also included in the training set. In contrast, Chen et al.~\cite{Chen2018JPCL} trained two deep NNs on 90,000 data points of two singlet states of CH$_2$NH. Again, data points were obtained from Born-Oppenheimer MD simulations and NAMD simulations starting from hopping geometries. 
In both studies, the NAMD simulations of the reference method could be successfully reproduced. 

Instead of approximating the hopping probability, the NACs can also be approximated from PESs, gradients and Hessians~\cite{Thiel1999JCP,Koeppel2001JCP,Koeppel2006MP,Maeda2010JCTC,Kammeraad2016JPCL,Gonon2017JCP}. We made use of this relation and the fact that ML Hessians can be computed efficiently, and carried out NAMD simulations with the surface hopping method for sulfur-dioxide, thioformaldehyde and the methylenimmonium cation~\cite{Westermayr2020JPCL}. 

\subsection*{NAMD simulations with ML energies, forces, and couplings}
In addition to energies and forces, SOCs need to be fitted with ML models when states of different spin multiplicities become relevant. Furthermore, when approximative schemes for the computation of hopping probabilities fail, the ML models need to learn NACs. One of the first studies, where NACs were fitted, used 1,000 and 10,000 data points to train kernel-ridge regression models to reproduce NAMD simulations of a one-dimensional system. However, especially in critical regions of the PESs, the ML models could not replace quantum chemical calculations and so 13-16\% electronic structure calculations were required during an NAMD simulation~\cite{Dral2018JPCL}. The authors highlighted this as a drawback, because efficient simulations should be performed purely with ML and should not rely on intermediate quantum chemical calculations. Moreover, each entry of the NAC vectors was fitted by a separate kernel-ridge regression model, which turned out to be insufficiently accurate. 
 
As indicated before, we also aimed for reproducing NAMD simulations with ML. We employed multi-layer feed-forward NNs trained on 4,000 data points of 3 singlet states of CH$_2$NH$_2^+$~\cite{Westermayr2019CS}. Short reference NAMD simulations based on electronic structure calculations could be reproduced. With the ML NAMD, long simulation times on the order of a nanosecond were successfully reached.
Significantly different from previous ML NAMD approaches is the smaller size of the training set required to reproduce NAMD simulations. Further, a multi-output ML model was used to fit all NAC vectors between different states of same spin multiplicity at once. We term such models multi-state models. Per definition, kernel-ridge regression, and similar approaches such as linear regression, are single-state models. In order to make multi-state predictions of such models possible, the energetic state has to be encoded explicitly by using for example an additional state kernel. This procedure enables to model several states simultaneously. We studied the use of multi-state descriptors with the QML toolkit~\cite{QML} for kernel-ridge regression models and showed that a multi-state description is generally superior to a single-state description in terms of accuracy~\cite{Westermayr2020MLST}. 

Lastly, we want to comment on the NACs as vectorial properties. It should be clarified that approaches relating a molecular input directly to NAC values do not provide rotational covariance. This drawback is independent of a single-state treatment, i.e., the use of a separate ML model for each coupling value, or a multi-state treatment, where all values are represented in one ML model. 
Very recently, Zhang et al. applied a symmetry-adapted high-dimensional NN ~\cite{Zhang2020JPCC} and treated the couplings as derivatives of NN representations. In this case, electronic friction was modelled via ML and applied for MD simulations of molecules at metal surfaces to treat the electron-nuclei coupling in a rotationally covariant manner. For the NAC vectors, we applied a similar strategy (similar also to force-only training for potentials), and implemented them as derivatives of virtual properties (i.e., non-existent in quantum chemistry) built by a deep NN~\cite{Westermayr2020JPCL}.

\subsection{Choosing the right descriptor}
Many of the aforementioned studies use kernel ridge regression models or NNs in combination with distance-based descriptors~\cite{Chen2018JPCL,Hu2018JPCL,Dral2018JPCL,Westermayr2019CSWang2018CPC,Zhang2018PRL} such as the matrix of inverse distances or the Coulomb matrix~\cite{Rupp2012PRL}.
It is worth mentioning that the accuracy of the ML PESs also depends on the type of descriptor. Molecular descriptors that represent atoms in their chemical and structural environment are often superior to those who treat complete molecules~\cite{Faber2018JCP,Christensen2019JCP,Christensen2020JCP}. The symmetry functions of Behler~\cite{Behler2007PRL,Behler2011JCP}, their weighted counterparts~\cite{Gastegger2018JCP,Herr2019JCP} or the FCHL (Faber-Christensen-Huang-Lilienfeld) representation~\cite{Faber2018JCP,Christensen2020JCP} work very well for NNs and the latter also for kernel-ridge regression and additionally provide permutation invariance. 

Further improvement can be provided by message passing neural networks~\cite{Gilmer2017}. Compared to hand-crafted molecular descriptors, the representation of molecules can be seen as a part of a deep NN and, thus, is generated automatically. For each training set, an accurate descriptor is intrinsically designed, which accounts for the chemical and structural environment of a molecule. Examples for such networks are SchNet~\cite{Schuett2018JCP,Schuett2019JCTC}, the DTNN~\cite{Schuett2017NC}, PhysNet~\cite{Unke2019JCTC}, or HIP-NN~\cite{Lubbers2018JCP}. For the excited states, the SchNarc~\cite{Westermayr2020JPCL} approach offers this type of descriptor.

\section{Conclusion and outlook}
To conclude, ML methods are powerful and can be used to speed up current MD approaches for the excited states. They have been successfully applied to circumvent existing problems due to the expenses of the underlying electronic structure methods.

While the fitting of diabatic potentials is generally more favorable, those methods are limited by the challenges that arise in finding meaningful diabatic potentials. Up to date, diabatization procedures are tedious and often not feasible for large and complex systems. ML models have been successfully applied to improve these processes~\cite{Williams2018JCP,Xie2018JCP,Guan2019PCCP}, but methods to treat large and complex polyatomic systems in the diabatic basis are still lacking. 

To investigate the photodynamics of polyatomic molecules, mixed quantum-classical MD simulations in the adiabatic basis thus often remain the method of choice. One advantage is that the direct output of a quantum chemical calculation is given in the adiabatic basis and so the obtained potential energies and forces can be directly fitted with an ML model. By applying approximations for the computation of transition probabilities from one state to another, the photodynamics can be studied efficiently with ML~\cite{Hu2018JPCL,Chen2018JPCL,Westermayr2020JPCL}. When approaches aim for additionally fitting the coupling values between different electronic states, inconsistencies in the data need to be considered carefully. Those have to be either removed from the training set or the training process itself has to be adapted in order to achieve successful training. Both approaches have been applied and were used for NAMD simulations~\cite{Dral2018JPCL,Westermayr2019CS,Westermayr2020JPCL}.

The current challenges that have to be tackled when replacing quantum chemical calculations in photodynamics simulations of large and complex polyatomic molecules are the efficient training set generation and the accurate fitting of a manifold of energetic states, forces, and couplings between them. Many recent ML approaches required several thousand of data points for small polyatomic molecules and energies, forces, and couplings were often trained in separate ML models, leading to unsatisfactory accuracy. The development of an ML model that can treat all properties for photodynamics simulations at once is clearly desirable. While current studies struggle with fitting the excited states of one molecule, the transferability of ML potentials for the excited states is far from being achieved. Studies that try to fit more molecules than just one could give first insights into the possibility of extrapolating throughout chemical compound space with ML also for the excited states.

\section*{Acknowledgements}
This work was financially supported by the Austrian Science Fund, W 1232 (MolTag) and the uni:docs program of the University of Vienna (J.W.). P. M. thanks the University of Vienna for continuous support, also in the frame of the research platform ViRAPID.
\section*{References}

\end{document}